\documentclass[allclo]{FBSart}

\usepackage{latexsym}
\usepackage{graphics}
\usepackage{latexsym}                            
\usepackage{amsfonts}                            
\usepackage{amssymb}                             
\usepackage{amsmath}                             
\usepackage[mathscr]{eucal}                      

\newcommand{\Dlr}{\stackrel{\leftrightarrow}{D}}

\title{
\begin{flushleft}
\vspace*{-2.5cm}
{\small DESY 04-205}\\[-0.5em]
{\small LU-ITP 2004/040}\\[-0.5em]
{\small Edinburgh 2004/26}
\vspace*{0.8cm}
\end{flushleft}
Generalized Parton Distributions in Full Lattice QCD
\footnote{Presented by Ph.~H\"agler at Light-Cone 2004, Amsterdam, 16 - 20 August}}
\author{M.~G\"ockeler$^{1,2}$, Ph.~H\"agler$^3$, R.~Horsley$^4$, D.~Pleiter$^5$, P.E.L.~Rakow$^6$,
A.~Sch\"afer$^2$, G.~Schierholz$^7$ and J.M.~Zanotti$^5$ (QCDSF collaboration)}

\institute{$^1$Institut f\"ur Theoretische Physik,
Universit\"at Leipzig, D-04109 Leipzig, Germany\newline
$^2$Institut f\"ur Theoretische Physik, Universit\"at Regensburg,
D-93040 Regensburg, Germany \newline
$^3$Department of Physics and Astronomy, Vrije Universiteit,
1081 HV Amsterdam, NL\newline
$^4$School of Physics, University of Edinburgh,
Edinburgh EH9 3JZ, UK\newline
$^5$John von Neumann-Institut f\"ur Computing 
NIC / DESY, D-15738 Zeuthen, Germany\newline
$^6$Theoretical Physics Division, Dep.~of
Math.~Sciences, University of Liverpool,\newline
\hspace*{2mm}Liverpool L69 3BX, UK\newline
$^7$Deutsches Elektronen-Synchrotron
DESY, D-22603 Hamburg, Germany}
\runningauthor{Ph.~H\"agler et al.}
\runningtitle{LC 2004}
\begin{document}
\maketitle
\begin{abstract}
We present recent results on generalized parton distributions from dynamical lattice QCD
calculations. Our set of twelve different combinations of couplings and quark masses allows
for a preliminary study of the pion mass dependence of the transverse nucleon structure.
\end{abstract}
\vspace{-1mm}
\section{Introduction}
\label{intro}
The investigation of the nucleon structure in the framework of QCD
is a major task in today's particle physics.
Generalized parton distributions (GPDs) have opened new ways of studying the
complex interplay of longitudinal momentum and transverse
coordinate space as well as spin and orbital angular momentum degrees of freedom
in the nucleon.
The twist-2 GPDs $H$, $E$, $\widetilde H$ and $\widetilde E$
of quarks in the nucleon are defined
by the following parameterization of off-forward nucleon matrix elements \vspace{-1mm}
\begin{multline}
  \!\!\!\! \left\langle P',\lambda ' \right|
  \int \frac{d \lambda}{4 \pi} e^{i \lambda x}
  \bar \psi (-\frac{\lambda}{2}n)\!
  \gamma^\mu
  {\cal U} \psi(\frac{\lambda}{2} n)
  \left| P,\lambda \right \rangle
  = \; \\
  \overline U(P',\lambda ') \left( \!\gamma^\mu  H(x, \xi, t)
 +\frac{i \sigma^{\mu \nu} \Delta_\nu} {2 m} E(x, \xi, t) \!\right)   U(P,\lambda)\,,
\label{VectorOp}
\end{multline}
for the helicity independent vector case, and
\begin{multline}
  \!\!\!\! \left\langle P',\lambda ' \right|
  \int \frac{d \lambda}{4 \pi} e^{i \lambda x}
  \bar \psi (-\frac{\lambda}{2}n)\!
  \gamma_5 \gamma^\mu
  {\cal U}\psi(\frac{\lambda}{2} n)
  \left| P,\lambda \right \rangle
  = \;  \\
  \overline U(P',\lambda ') \left(\! \gamma_5 \gamma^\mu  \widetilde H(x, \xi, t)
 + \frac{\gamma_5 \Delta^\mu} {2 m} \widetilde E(x, \xi, t) \!\right)   U(P,\lambda) \,,
\label{AxialVectorOp}
\end{multline}
for the helicity dependent axial vector case \cite{Muller:1998fv,Ji:1996nm,Radyushkin:1997ki}.
Wilson lines ensuring gauge invariance of the bilocal operators are denoted by $\cal{U}$.
Here and in the following we do not show explicitly the dependence of the GPDs on the
resolution scale $Q^2$.
\begin{figure}
\begin{center}
\resizebox{0.60\textwidth}{!}
{
\rotatebox{270}{
  \includegraphics{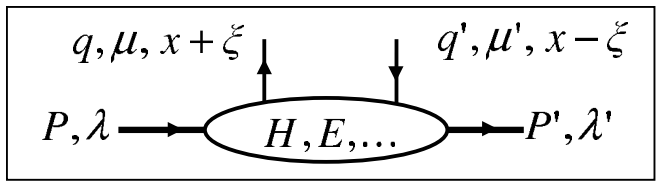}
}      }
\caption{Graphical representation of GPDs as part of a scattering amplitude}
\label{fig:GPDs}       
\vspace{-0.5cm}
\end{center}
\end{figure}
Fig.(\ref{fig:GPDs}) shows the definition of the particle momenta and helicities.
The momentum transfer (squared) is given by $\Delta=P'-P$ ($t=\Delta^2$).
Using the light cone vector $n$ we define the longitudinal momentum transfer by $\xi=-n\cdot \Delta/2$.
The proper definition of the twist-2 tensor or quark helicity flip GPDs
$H_T$, $E_T$, $\widetilde H_T$ and $\widetilde E_T$ can be found in \cite{Diehl:2001pm}.
GPDs provide a solid framework in QCD to relate many different aspects of hadron pyhsics:
\begin{itemize}
\item The forward limit $\Delta \to 0$ of certain GPDs reproduces the well known parton distributions,
that is $H(x, \xi=0, t=0)=q(x)=f_1(x)$, $\widetilde H(x, 0, 0)=\Delta q(x)=g_1(x)$
and $H_T(x, 0, 0)=\delta q(x)=h_1(x)$.
\item The integral over the longitudinal momentum fraction $\int dx$ of the GPDs gives the
Dirac, Pauli, axial, pseudo-scalar, tensor etc. form factors, 
$\int dx H(x, \xi, t)=F_1(t)$, $\int dx E=F_2(t)$,
$\int dx \widetilde H=g_A(t)$, $\int dx \widetilde E=g_P(t)$,
$\int dx H_T=g_T(t)$
etc.
\item The Fourier transforms $\int d\Delta_\perp e^{-i b_\perp \cdot \Delta_\perp}$
of the GPDs $H$, $\widetilde H$ and $H_T$ at $\xi=0$
are coordinate space probability densities in the impact parameter $b_\perp$ \cite{Burkardt:2000za}.
\item The forward limit of the $x$-moment of the GPD $E$, $\int dx x E(x,0,0)=B_{20}(0)$,
allows for the determination of the quark orbital angular momentum contribution to the nucleon spin,
$L^q=1/2 (\langle x\rangle +B_{20}-\Delta q)$, where $\langle x\rangle$ is the quark momentum fraction.
\end{itemize}
The fact that the impact parameter dependent quark distributions like
\begin{eqnarray}
q(x,b_\perp)
\equiv\int d^2\Delta_\perp e^{-i b_\perp \cdot \Delta_\perp} H(x,\xi=0,t=-\Delta_\perp^2)
\end{eqnarray}
have the interpretation of probability densities in the transverse plane
marks a major step forward in the understanding of the nucleon structure.
A very interesting observation in this respect is that at large longitudinal momentum fraction, $x\to 1$,
i.e. when the active parton carries the whole nucleon momentum and the
spectators give only a negligible contribution, the active parton becomes
the center of momentum of the nucleon at $R_\perp=0$. Accordingly, the impact parameter
profile of the nucleon will be strongly peaked at $b_\perp=0$. Keeping in mind
that the parton distributions eventually vanish in the limit $x\to 1$, we expect that
$\lim_{x \to 1}q(x,b_{\perp })/q(x) \varpropto \delta ^{2}\left( b_{\perp}\right)$.
Transferred to momentum space, we have the striking prediction that the normalized GPD $H(x,0,t)/q(x)$ will
be constant in the limit $x\to 1$\cite{Burkardt:2002hr,Burkardt:2004bv}.

Interesting recent results on GPDs using available 
experimental data on form factors and PDFs can be found in \cite{Diehl:2004cx}.

On the lattice side, we cannot deal directly with matrix elements
of bilocal light-cone operators. Therefore we first transform the 
LHS in Eqs.(\ref{VectorOp},\ref{AxialVectorOp})
to Mellin space by integrating over $x$, i.e. $\int dx x^{n-1}$. This results in matrix elements
of towers of local operators which are in turn parameterized in terms of so-called generalized
form factors (GFFs) $A(t),B(t)\ldots$,
\begin{eqnarray}
\!\!\left\langle P^{\prime },\lambda '\right| \bar{\psi}(0)
    \Gamma i\Dlr^{\{\mu _{1}}\!\!\cdots i\Dlr^{\mu _{n}\}}\psi (0)\left| P, \lambda \right\rangle
 &=&\overline{U}(P',\lambda ')
 \left(a_{\Gamma }^{\mu_{1}\ldots \mu _{n}}A(t)\right. \nonumber \\
&+& \left. b_{\Gamma }^{\mu_{1}\ldots \mu _{n}}B(t)+\cdots \right)U(P,\lambda),
\label{ME1}
\end{eqnarray}
where the subtraction of traces is implicit.
The explicit parameterization for the tower of vector operators, $\Gamma\equiv\gamma^\mu$, is
given in \cite{Hoodbhoy:1998yb} in
terms of the GFFs $A_{ni}(t),B_{ni}(t)$ and $C_{n0}(t)$.
In the axial vector case, the corresponding expression in terms of the GFFs
$\widetilde A_{ni}(t)$ and $\widetilde B_{ni}(t)$ is shown in \cite{Diehl:2003ny}. 
Parameterizations for the tensor GPDs have been derived recently in \cite{Hagler:2004yt,Chen:2004cg}.
Having computed the GFFs in e.g. lattice QCD, it is fairly simple to get the moments of the
corresponding GPDs by using the following polynomial relations \cite{Ji:1997gm}
\begin{eqnarray}
H^{n}(\xi ,t) \!\!&\equiv& \!\!\int\limits_{-1}^{1}dxx^{n-1}H(x,\xi ,t)
=\!\!\sum_{i=0, \text{even}}^{n-1}\left( -2\xi \right) ^{i}A_{ni}(t)
+\left. \left( -2\xi \right) ^{n}C_{n0}(t)\right| _{n
\text{ even}},\,  \nonumber \\
E^{n}(\xi ,t) \!\!&=&\!\!\sum_{i=0,
 \text{even}}^{n-1}\left( -2\xi
\right) ^{i}B_{ni}(t)
-\left. \left( -2\xi \right)
^{n}C_{n0}(t)\right| _{n\text{ even}},\, \nonumber\\
\widetilde H^{n}(\xi ,t) \!\!&=&\!\!\sum_{i=0, \text{even}}^{n-1}\left( -2\xi \right) ^{i}\widetilde A_{ni}(t),\,\,\,\,
\,\,\,\,\,\,
\widetilde E^{n}(\xi ,t) =\!\!\sum_{i=0, \text{even}}^{n-1}\left( -2\xi \right) ^{i}\widetilde B_{ni}(t) .  \label{invv}
\end{eqnarray}
In the limit $\xi\to 0$, the complete nucleon structure related to $H$ and $\widetilde H$ is according
to Eq.(\ref{invv}) represented by the sets of GFFs $A_{n0}(t)$ and $\widetilde A_{n0}(t)$, $n=1\ldots\infty$.
%
\begin{figure}[t]
\begin{center}
\resizebox{0.95\textwidth}{!}
{
\rotatebox{270}{
  \includegraphics{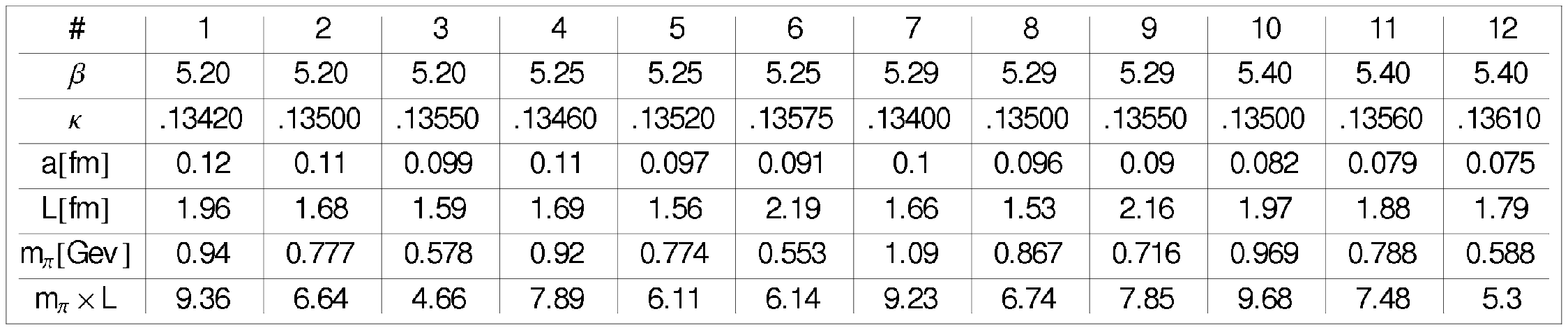}
}      }
\caption{Some parameters of our calculation.}
\label{fig:Para}       
\vspace{-0.5cm}
\end{center}
\end{figure}
\section{Lattice results}
The lattice calculation and extraction of the GFFs from nucleon
two- and three-point-functions follows the standard procedure as outlined in e.g. 
\cite{Hagler:2003jd,lattice3,Gockeler:2004vx}.
Our dynamical computations are currently based on a set of coupling- and quark mass-combinations
with up to $O(1200)$ configurations and with pion masses ranging from $1.1$ down to $0.55$ GeV.
The lattice spacing lies in between $\approx 0.07$ and $\approx 0.12$ fm.
Since we are using non-perturbatively $O(a)$ improved Wilson fermions, we do not expect strong
discretization effects \cite{Gockeler:2004vx}.
A list of the parameters is given in Fig.(\ref{fig:Para}).
We do not take into account the disconnected contributions which are numerically much
harder to evaluate. This is irrelevant as long as we concentrate on the iso-vector channel $u-d$ (as we do here),
for which the disconnected pieces would cancel due to iso-spin symmetry.
\begin{figure}[t]
\begin{center}
\resizebox{1.\textwidth}{!}
{
\rotatebox{270}{
  \includegraphics{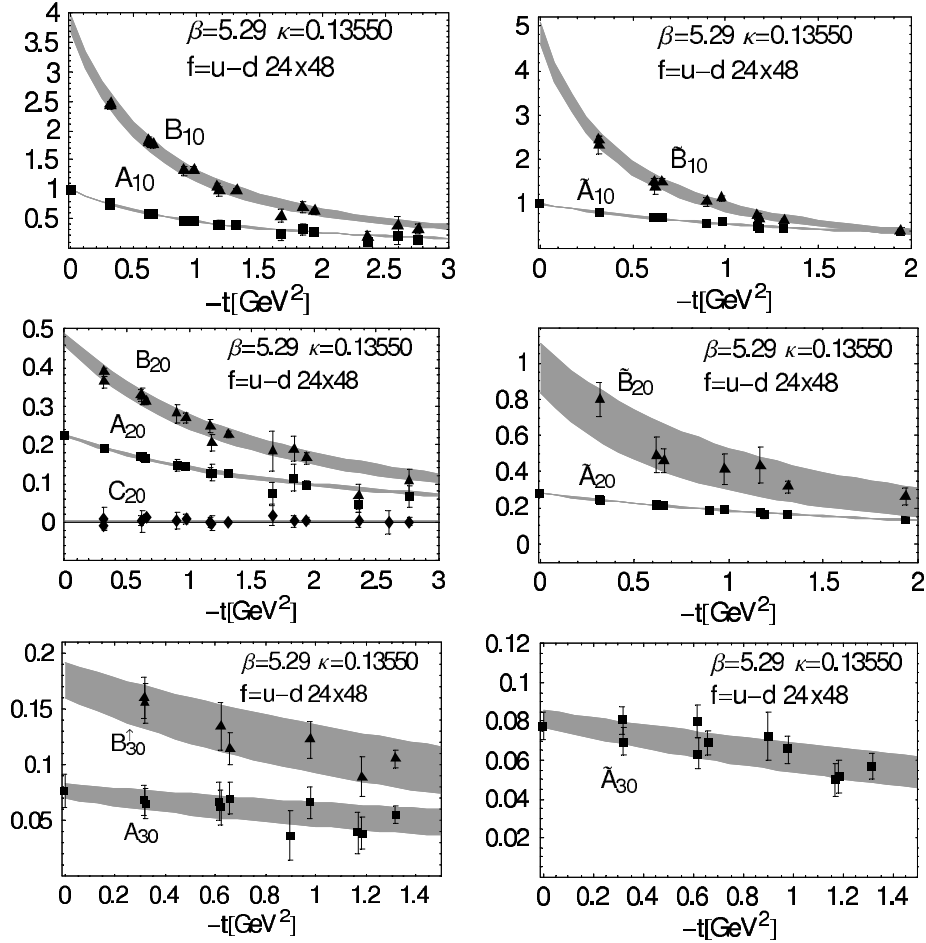}
}      }
\caption{Lowest three moments of the GPDs $H(x,0,t)$, $E(x,0,t)$ (left column) and $\widetilde H(x,0,t)$, $\widetilde E(x,0,t)$
(right column).}
\label{fig:First13550}       
\vspace{-0.5cm}
\end{center}
\end{figure}
To give an overview of our results we show in Fig.(\ref{fig:First13550})
the lowest three moments of the GPDs $H(x,0,t)$, $E(x,0,t)$ and $\widetilde H(x,0,t)$ as
well as the lowest two moments of $\widetilde E(x,0,t)$. 
Up to now, only the lowest moments $n=1$ which we show here are
non-perturbatively renormalized, and operator mixing has to be taken into account
for a proper renormalization of the $x^2$-moments for $i>0$ \cite{Gockeler:2004xb}.
We do not expect the renormalization to affect the following general discussion
and the extraction of the dipole masses below.
Please note that for reasons of clarity $B_{30}(t)$ has been
shifted upwards by $0.05$, as indicated by the superscript arrow
in Fig.(\ref{fig:First13550}), lower left panel.
All GFFs have been fitted using a standard dipole ansatz,
\begin{equation}
A_{ni}^\text{dipole} (t) = \frac{A_{ni}^\text{dipole}(0)}{\left( 1 - {t/m_D^2}
  \right)^2} \ ,
\label{dipole}
\end{equation}
except for the pseudo-scalar FF $\widetilde B_{10}(t)=g_P(t)$, where we used a dipole
ansatz including the pion-pole,
\begin{equation}
\widetilde B_{10}^{\pi\text{pole}}(t) = \frac{\widetilde B_{10}^{\pi\text{pole}}(0)}
{\left( 1 - {t/m_D^2}\right)^2\left( 1 - {t/m_\pi^2}\right)}\,.
\label{pipole}
\end{equation}
The fits given by the gray bands in Figs.(\ref{fig:First13550})
show an overall satisfactory description of the lattice numbers. While the statistical errors
for the lowest two moments are rather small, they tend to be somewhat larger for the $x^2$-moment.
The statistics are particularly poor for $A_{32}$, $\widetilde A_{32}$ and the $\widetilde B_{3i}$ GFFs, which
are not shown in this work.
It turns out that the moments of $H$ and $\widetilde H$ in the iso-vector channel are
a factor of $2\ldots 5$ smaller than the corresponding moments of $E$ and $\widetilde E$, while $C_{20}$
which contributes to both $H$ and $E$ is compatible with zero. Our results indicate furthermore
strong cancellations of $u$ and
$d$ flavor contributions in the iso-singlet channel for
the helicity independent GPD $E^{u+d}$.
\begin{figure}
\begin{center}
\resizebox{0.95\textwidth}{!}
{
\rotatebox{270}{
  \includegraphics{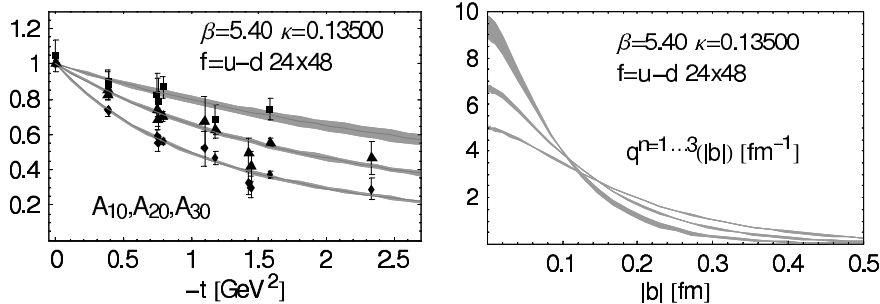}
}      }
\caption{Change of slope of the GFFs (left)and narrowing of the corresponding $b_\perp$ dependent GFFs (right);
$n=1$ corresponds to the lowest (widest) curve in the left(right)-hand plot.}
\label{fig:Impact}       
\end{center}
\end{figure}
Let us now take a look at the $t$-dependence of the GPDs
and the impact parameter dependent quark distribution $q(x,b_\perp)$.
The observation from section (\ref{intro}) that $\lim_{x\to 1}H(x,0,t)/q(x)=1$
translates directly to the statement that the GFFs $A_{n0}(t)$ become constant for very large $n$
since increasingly more weight lies on values of $x$ close to one for $\lim_{n\to \infty}$.
To be precise, we expect that $\lim_{n\to\infty} A_{n0}(t)/A_{n0}(0)=1$, independent of $t$.
Since the values for the momentum transfer squared in our computation are unevenly
distributed in the interval $-3.5<t<0$ GeV,
and since there is in particular a gap between $t=0$ and the smallest non-zero $t$ which can
be realized on the given lattices, $t_\text{min}$, it is not useful to perform a discrete
Fourier transformation (FT) of our results to impact parameter space.
Instead, we directly use the continuous FT of the dipole ansatz Eq.(\ref{dipole}).
The normalized impact parameter dependent moments $q^{n=1\ldots 3}(b_\perp)$ with
$\int db_\perp q^{n=1\ldots 3}(b_\perp)=1$ are shown in Fig.(\ref{fig:Impact}).
For reasons of comparison we show in Fig.(\ref{fig:Impact}) also the corresponding
GFFs, normalized to $A_{n0}^\text{dipole} (0)=1$. The gray bands in the impact parameter plot
reflect the errors associated with the dipole masses, $\Delta m_D$.
The results in Fig.(\ref{fig:Impact}) show very nicely the anticipated
flattening of the GFFs and the narrowing of the corresponding
distributions in impact parameter space, going from $n=1$ to $n=3$.
This flattening of the GPDs has first been observed in \cite{Hagler:2003is}.

Since we have results for a large number of pion masses at hand, it is mandatory to
concentrate in the following on the important issue of the $m_\pi$ dependence
and the chiral extrapolation of our calculation. Due to the use of Wilson fermions,
the dominant part of our computation is for pion masses above $600$ MeV where chiral perturbation
theory in its current form is most probably not applicable. In particular for the
dipole masses $m_D$, we therefore resort to a purely phenomenologically motivated
pion mass dependence of the form $m_D=m_D^0+\alpha m_\pi^2$. We have checked that
a linear dependence on $m_\pi$ is clearly unfavored by our results.
\begin{figure}
\begin{center}
\resizebox{1.\textwidth}{!}
{
\rotatebox{270}{
  \includegraphics{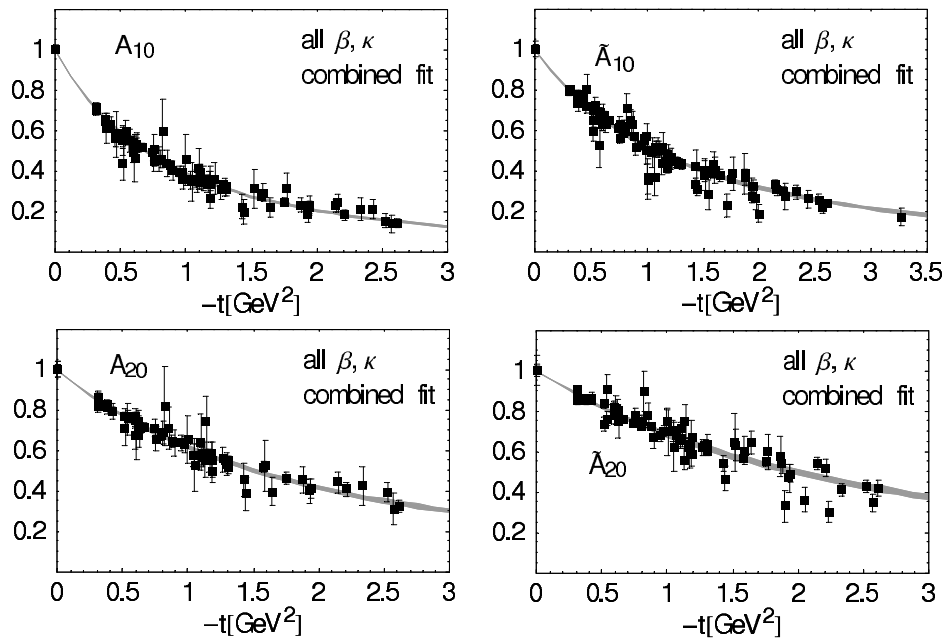}
}      }
\caption{Joined results for all $\beta$, $\kappa$ combinations for the lowest two moments of $H(x,0,t)$
(left column) and $\widetilde H(x,0,t)$ (right column).}
\label{fig:AllData1}       
\vspace{-0.5cm}
\end{center}
\end{figure}
For the following analysis we have normalized our results to $A_{n0}^\text{dipole}(0)=1$ first.
One option is then to perform dipole fits for all individual $\beta$, $\kappa$ combinations for a given
moment $n$ and fit the pion mass dependence of the dipole masses with the given ansatz.
However to optimally utilize the information on the dipole masses from
our collection of results, it is better to use a combined dipole-mass/pion-mass fit
of the simple form
\begin{equation}
\bar A_{n0}^{\text{dipole},m_\pi} (t) = \frac{1}{\left( 1 - {t/(m_D^0+\alpha m_\pi^2)^2}
  \right)^2} \ ,
\label{chiraldipole}
\end{equation}
depending on the two parameters $m_D^0$ and $\alpha$.
Having done the fit, we may shift the raw numbers
to a common curve given by Eq.(\ref{chiraldipole}) at the physical pion mass, $m_{\pi,\text{phys}}=139$ MeV.
The results of this procedure are shown in Fig.(\ref{fig:AllData1})
for the GFFs $A_{n0}$ and $\widetilde A_{n0}$, $n=1,2$. The fit parameters are found to be
\begin{eqnarray}
A_{10}\, &:& \, m_D^0 = 1.27\!\pm\! .01 \text{ GeV}, \,\,\, \alpha= 0.29\!\pm\! .02 \text{ GeV}^{-2}\nonumber\\
A_{20}\, &:& \, m_D^0 = 1.90\!\pm\! .03 \text{ GeV}, \,\,\, \alpha= 0.14\!\pm\! .05 \text{ GeV}^{-2}\nonumber\\
\widetilde A_{10}\, &:& \, m_D^0 = 1.59\!\pm\! .02 \text{ GeV}, \,\,\, \alpha= 0.16\!\pm\! .03 \text{ GeV}^{-2}\nonumber\\
\widetilde A_{20}\, &:& \, m_D^0 = 2.18\!\pm\! .05 \text{ GeV}, \,\,\, \alpha= -0.10\!\pm\! .06 \text{ GeV}^{-2}\, .
\label{FitRes}
\end{eqnarray}
Although the corresponding rather large error does not allow for a definite conclusion, it is interesting
to note that the slope $\alpha$ changes sign going from $\widetilde A_{10}$ to $\widetilde A_{20}$.
We observe an overall nice clustering of the lattice points in Fig.(\ref{fig:AllData1})
much in favor of our
ansatz Eq.(\ref{chiraldipole}).
The actual fits are given by the gray bands and show
a very good $\chi^2/\text{NODOF}$ lying below $0.2$ for the helicity independent and below $0.5$ for
the helicity dependent case.
In particular for $\widetilde A_{n0}$, a small fraction of the
lattice points seem to be significantly below the
main line. Future investigations have to show if this is due to an unaccounted dependence on the lattice
spacing, the (finite) lattice volume or just insufficient statistics.
An attempt to compare our results for the axial and tensor coupling, $g_A(0)$
and $g_T(0)$, with chiral
perturbation theory is described in \cite{Khan:2004vw}.
We plan to extract the quark orbital angular momentum contribution to the
nucleon spin as soon as the missing renormalization constants
become available.
\section{Conclusion}
We have computed the lowest moments of GPDs for a number of different pion masses
in the range above $550$ MeV. The results are very well described by a
combined dipole-mass/pion-mass fit.
While we observe no sign of chiral logarithms in the available range of pion masses,
let us note that first promising lattice calculations towards the chiral region are
currently under way using overlap \cite{Gurtler:2004ac} as well as domain wall
fermions \cite{Renner:2004ck}.

\subsection*{Acknowledgments}
The numerical calculations have been performed on the Hitachi SR8000 at LRZ (Munich), 
on the Cray T3E at EPCC (Edinburgh) and on the APEmille at NIC/DESY (Zeuthen). 
This work has been supported in part by the DFG (Forschergruppe Gitter-Hadronen-Phaenomenologie),
the EU Integrated Infrastructure Initiative ``Hadron Physics" as well as
``Study of Strongly Interactive Matter" (in parts under contract number RII3-CT-2004-506078).

%
%

%
\end{document}